\newif\ifpreprint
\preprinttrue

\ifpreprint
\documentclass[nofootinbib,aps,prd,preprint,12pt]{revtex4}
\else
\documentclass[aps,prl,twocolumn]{revtex4}
\fi

\usepackage{amsmath,amsfonts}
\usepackage{cancel}

\newcommand{\beq}{\begin{equation}}
\newcommand{\eeq}{\end{equation}}
\newcommand{\beqa}{\begin{eqnarray}}
\newcommand{\eeqa}{\end{eqnarray}}
\newcommand{\nn}{\nonumber  \\}

\begin{document}

\title{A Grassmannian \'Etude in NMHV Minors}

\author{Dhritiman Nandan, Anastasia Volovich and Congkao Wen}

\affiliation{Brown University, Providence, Rhode Island 02912, USA}

\begin{abstract}
Arkani-Hamed, Cachazo, Cheung and Kaplan have proposed a Grassmannian
formulation for the S-matrix of ${\cal N}=4$ Yang-Mills as an
integral over link variables. In parallel work, the connected prescription for
computing tree amplitudes in Witten's twistor string theory has also been
written in terms of link variables. In this paper we extend
the six- and seven-point results of arXiv:0909.0229 and
arXiv:0909.0499 by providing a simple analytic proof of the
equivalence between the two formulas for all tree-level NMHV
superamplitudes. Also we note that a simple deformation
of the connected prescription integrand gives 
directly the ACCK Grassmannian integrand in the limit when 
the deformation parameters equal zero.

\end{abstract}

\maketitle

\section{Introduction}
The twistor string theory formulation of Yang-Mills scattering amplitudes by Witten \cite{Witten:2003nn} has been a great step forward in unearthing a host of properties of scattering amplitudes, hitherto unseen via the standard methods of quantum field theory.  
A connected prescription formula for computing all tree level
superamplitudes in twistor string theory has been written down by Roiban, Spradlin and one of the 
authors in~\cite{RSV}, based on Witten's
proposal that the N${}^{k-2}$MHV superamplitude should be given by
the integral of an open string current algebra correlator over the
space of degree $k-1$ curves in supertwistor space
${\mathbb{P}}^{3|4}$. As noted in~\cite{RSV} an essential
feature of the connected prescription is that the resulting integral
for any physical space amplitude completely localizes, allowing it
to be expressed as a sum over roots or equivalently as a contour
integral (see also \cite{Vergu:2006np}).
Recently a ``linked'' version of the formula had been written
in \cite{Spradlin:2009qr} and \cite{Dolan:2009wf} by reformulating the
original connected prescription amplitude in terms of the link
variables introduced in \cite{ArkaniHamed:2009si}. A remarkable new
contour integral over a Grassmannian of these link variables,
which apparently encapsulates information about leading
singularities of ${\cal N}=4$ Yang-Mills loop amplitudes in addition to tree-level information,
has been written down by Arkani-Hamed, Cachazo, Cheung and Kaplan (ACCK) in
\cite{ArkaniHamed:2009dn}. See also
\cite{Mason:2009qx}\cite{ArkaniHamed:2009vw}\cite{Bullimore:2009cb}\cite{Kaplan:2009mh}\cite{ArkaniHamed:2009sx}
for related recent developments.

It has been proven for the case of six and seven particles
\cite{Spradlin:2009qr}\cite{Dolan:2009wf} that the residues of
both the linked-connected formula and the ACCK formula compute BCFW
representations \cite{Britto:2004ap}\cite{Britto:2005fq} of tree amplitudes. In this paper we  make the
connection between the linked-connected prescription formula from
twistor string theory and the ACCK proposal more transparent by offering
a simple analytic proof between the two formulas for all tree-level NMHV
superamplitudes.
Also we note that a simple deformation
of the connected prescription integrand by non-zero parameters gives 
directly the Grassmannian integrand in the limit when 
the deformation parameters equal zero.
Specifically, the ACCK Grassmannian integrand arises from the linked-connected
formula in a simple limit when the second terms in all
sextic polynomials are zero (see formula (\ref{deformation})).

In section II we review some of the recent developments and write
down a general formula (\ref{nmhvformula}) for $n$-point NMHV amplitudes in
terms of minors in a convenient way. 
In section III we show how to get the BCFW contours
from the linked-connected prescription for the six and seven point
NMHV amplitudes in a simple way, followed by the general proof for all
$n$-point NMHV amplitude by using the global residue theorem (GRT).
In the appendix we present the ten--point case as a concrete example.

\section{Review of Recent Developments}
\subsection{Review of Dual S-Matrix Formulation} Recently Arkani-Hamed, Cachazo, Cheung and Kaplan \cite{ArkaniHamed:2009dn} have
conjectured a formula for a dual formulation for the S-Matrix of
$\mathcal{N}=4$ SYM. According to their proposal the planar,
color stripped, $n$ particle, N${}^{k-2}$MHV amplitudes are associated
with contour integrals over a Grassmannian
\beqa
 {\cal L}_{n;k}({\cal W}_a) =\frac{1}{{\rm Vol}(GL(k))} \int \frac{d^{k \times n} C_{\alpha
a}}{(12\cdots k) \, (23\cdots (k+1) \,) \, \cdots (n 1 \cdots (k-1)
\,)} \prod_{\alpha = 1}^k \delta^{4|4}(C_{\alpha a} {\cal W}_a)
\eeqa
where  the ${\cal W}_a$ are twistor variables  obtained by
Fourier transforming with respect to the $\lambda_a:$ ${\cal W} =
(W | \tilde \eta) = (\tilde \mu, \tilde \lambda| \tilde \eta),$ and
\beqa
 (m_1  \cdots m_k) \equiv \epsilon^{\alpha_1 \cdots \alpha_k}
C_{\alpha_1 m_1} \cdots C_{\alpha_k m_k}.
 \eeqa
Here, $C_{\alpha a}$ is a $k \times n$ matrix and its `minor', $(m_1
\cdots m_k)$ is the determinant of the $k \times k$ submatrix made by
only keeping the $k$ columns $m_1, \cdots, m_k$. The integrand of
this formula has a $GL(k)$ symmetry  under which $C_{\alpha a} \to
L_{\alpha}^\beta C_{\beta a}$ for any $k \times k$ matrix $L$, and
so one has to gauge fix  by dividing by Vol(GL$(k)$). This formula
has manifest cyclic, parity, superconformal and also dual
superconformal symmetry \cite{ArkaniHamed:2009vw}.

The outstanding feature of this formula
is that, interpreting the integral as a multidimensonal contour
integral in momentum space, the residues of the integrand give a
basis for obtaining tree level amplitudes as well as all loop
leading singularities.
\subsection{NMHV tree amplitude from ACCK}
A general formula for determining which residues correspond to  tree
amplitudes for the $n$ particle NMHV case has been given in \cite{ArkaniHamed:2009dn} which we will now review.
Following their notation we denote
a residue when $n-5$ minors $(i_1~i_1+1~i_1+2), \dots,
(i_{n-5}~i_{n-5}+1~i_{n-5}+2)\rightarrow 0$ as
$\{i_1,i_2,\cdots,i_{n-5}\}$, and it is antisymmetric. Then NMHV tree amplitude is given
by the sum of residues
\begin{equation}
\begin{aligned}\label{Nimabcf}
A_{n,\rm{BCFW}}^{\rm {NMHV}}&=(-1)^{n-5}\underbrace{\mathcal{O}\star
\mathcal{E} \star \mathcal{O} \star \mathcal{E} \dots}\\ &~~~~ ~~~~~~~~~~(n-5) \,
{\rm factors}
\end{aligned}
\end{equation}
where $\mathcal{O}$ is the set of odd numbered particles and
$\mathcal{E}$ is the set of even numbered particles
\begin{equation}
\begin{aligned}
{\cal O} = \sum_{k \, {\rm odd}} \{k\}, \, ~~~{\cal E} = \sum_{k \,
{\rm even}} \{k\}
\end{aligned}
\end{equation}
and
\begin{equation}
\{i_1\} \star \{i_2\} = \begin{cases} \{i_1, i_2\}  & {\rm if}~ i_1 < i_2 \cr 0 & {\rm
otherwise} \end{cases}
\end{equation}
The above proposal can  also be motivated from the geometric picture presented in the recent papers  \cite{Korchemsky:2009jv} and \cite{Bullimore:2009cb}.

To get P(BCFW) (parity-conjugated BCFW terms) from BCFW, one can simply apply the GRT. 
For example, the BCFW terms of the seven-point NMHV amplitude can be written as
\begin{equation}
\begin{aligned}
A_7=\{1,2\}+\{1,4\}+\{1,6\}+\{3,4\}+\{3,6\}+\{5,6\}.
\end{aligned}
\end{equation}

\subsection{Review of the Linked-Connected Prescription}
Let us begin by reviewing some details of the connected prescription
formula \cite{RSV}. The $4|4$ component homogeneous coordinates for the $i$-th
particle in ${\mathbb{P}}^{3|4}$ are ${\cal Z}_i =
(\lambda_i^\alpha, \mu_i^{\dot{\alpha}}, \eta_i^A)$ with
$\alpha,\dot{\alpha} = 1,2$ and $A =1,2,3,4$. The connected formula
can be written explicitly in the following form:
\begin{equation}
\label{eq:integralone}
{\cal A}({\cal Z}) = \int \frac{d^{4k|4k} {\cal A}\, d^n \sigma\,
d^n \xi}{{\rm vol}\,GL(2)}
\prod_{i=1}^n \frac{ \delta^{4|4}({\cal Z}_i -
\xi_i {\cal P}(\sigma_i))}{\xi_i(\sigma_i-\sigma_{i+1})},
\end{equation}
where ${\cal P}$ is the degree $k-1$ polynomial given in terms of its $k$
${\mathbb{C}}^{4|4}$-valued supercoefficients ${\cal A}_d$ by
\begin{equation}
{\cal P}(\sigma) = \sum_{d=0}^{k-1} {\cal A}_d \sigma^d.
\end{equation}
As emphasized in~\cite{RSV} (see also~\cite{Vergu:2006np})
the integral~(\ref{eq:integralone}) must be interpreted as a contour
integral in a multidimensional complex space.  The delta functions
specify the contour of integration (specifically they indicate which poles to
include in the sum over residues). There is also a $GL(2)$
invariance, of the integrand and the measure, which needs to be gauged.
Taking the above connected prescription as a starting point
and motivated by~\cite{ArkaniHamed:2009si} one can express the
connected prescription~(\ref{eq:integralone}) into the form of
so-called link representation \cite{Spradlin:2009qr}, \cite{Dolan:2009wf}.

One can obtain the physical space amplitude from the link representation 
\begin{equation}
\label{eq:link2} {\cal A}(\lambda, \widetilde{\lambda}) =J \delta(\sum p_i) \oint d \tau\
U(c_{Ji}(\tau_{\gamma})),
\end{equation}
where the Jacobian $J$ generally depends on the parameterization of $c_{Ji}(\tau_{\gamma})$.
A general form of $U(c_{Ji})$ has been explicitly evaluated by Dolan and Goddard in
\cite{Dolan:2009wf}. For an amplitude with helicities $(\epsilon_1, \dots,
\epsilon_n)$ comprising $p$ strings with $\epsilon_{\alpha}=+$ and
$p$ strings with $\epsilon_{\beta}=-$, their
explicit form is
\begin{equation} \label{dg1}
U(c)= F(c) \prod_{k,t} {1 \over S_{kt}},
\end{equation}
where $S_{kt}$ is the sextic
$S_{IJk:RSt}=c_{IS}c_{kt}c_{Jk:RS}c_{IJ:tR}-c_{It}c_{kS}c_{Jk:tR}c_{IJ:RS}$
with $c_{ij:rs}=c_{ir}c_{js}-c_{jr}c_{is}$, and
\begin{equation}
F(c)=\left(c_{IJ:RS}\right)^{N_R-p+2}c_{IR}^{p-3}c_{IS}^{p-3}
c_{JR}^{p-3}c_{JS}^{p-3} \prod_{
t\in\mathcal{P}'}c_{It}^{l-3}c_{Jt}^{l-3}
\prod_{k\in\mathcal{N}'}c_{kR}^{m-3}c_{kS}^{m-3}\prod_{k\in\mathcal{N}\atop
t\in\mathcal{P}} {1\over c_{kt}}\prod_{\alpha=1}^n d_{\alpha
,\alpha+1}, \label{dg2}
\end{equation}
where
$$d_{ir}=c_{ir},\quad d_{ri}=c_{ir},\quad d_{ij}={c_{iR}c_{jS}c_{jR}c_{iS}
\over c_{iR}c_{jS}-c_{jR}c_{iS}},\quad d_{rs}={
c_{Ir}c_{Js}c_{Is}c_{Jr}\over c_{Ir}c_{Js}-c_{Is}c_{Jr}},\quad
i,j\in\mathcal{N},\; r,s\in\mathcal{P}.$$
 We denote  $\mathcal{P}$ as the set of positive
helicity particles and $\mathcal{N}$ as the set of negative helicity
particles, and $N_R$ is the number of independent sextics, $l$ is the number of the negative helicity
particles, $m$ the number of the positive helicity
particles and $n=m+p$ is the total number of particles.\footnote{Here we exchange the helicities
$+\leftrightarrow-$, at the same time $c_{ij} \rightarrow c_{ji}$ with respect to \cite{Dolan:2009wf}.}

  \subsection{NMHV tree amplitude from the connected prescription}
In order to make the connection between the linked-connected  and ACCK
formulas more transparent, in this section we will express the
linked-connected formula in terms of  minors as in
the ACCK approach.\footnote{
We are grateful to Freddy Cachazo for encouraging us to
rewrite everything in terms of minors.
There are many different ways to write the formulas, but we
will pick the one which makes the proof simpler and has many other
nice properties as we will discuss later.}

Let us start with
helicity $(-+-+-++\dots ++)$, and take $I=1, J=3, R=2, S=4$, then
formula (\ref{dg1}) becomes
\begin{equation}
\begin{aligned}\label{npt}
U(c)=
(c_{52}c_{54}c_{13:24})^{n-6}(c_{12}c_{32}c_{34}c_{54}c_{56}c_{1n})
\prod^{n-1}_{\alpha=6}{c_{1
\alpha}c_{1,\alpha+1}c_{3\alpha}c_{3,\alpha+1} \over
c_{13:\alpha,\alpha+1}} \prod_{k\in\mathcal{P},t\in\mathcal{N}} {1
\over c_{kt}} \prod_{i=6}^n \frac{1}{S_{135:24i}}.
\end{aligned}
\end{equation}

Using the identity
\begin{equation}
\begin{aligned}\label{identity}
\delta(S_{ijk:rst})\delta(S_{ijk:rst'})=\delta(S_{ijk:rst})\delta(S_{ijk:rt't}){c_{it}c_{jk:rt}
\over c_{is}c_{jk:rs}},
\end{aligned}
\end{equation}
we can  transform the
sextics $S_{135:24i}$ in (\ref{npt}) to $S_{135:246}, S_{135:2,n-1,n}$, and 
$S_{135:i-1,i,i+1}$ to arrive at
\begin{equation}
\begin{aligned}\label{npt2}
U'(c)= {c_{35:26} c_{12} c_{13:n-1,n} c_{5,n-1} \prod^{n}_{\alpha=8}c_{5\alpha} 
\prod^{n-1}_{\beta=7}c_{3\beta} \prod^{n-2}_{\gamma=6}c_{1\gamma}
\over c_{52} c_{14} c_{13:67}c_{35:n-1,n} } {1 \over S_1S_2 \dots S_{n-5}}.
\end{aligned}
\end{equation}
We then translate it into minors, the result is\footnote{When $n=6$ or $n=7$
the minor $(567)$ does not appear in the denominator. And we put the minor $(135)$ in the numerator by hand to make the scale right, since $(135)=1$ for the helitiy we started.}
\beqa 
\label{nmhvformula}
A_n =
\frac{\mathcal{N}}{(123)(345)(567)(n-1~n~1)} {1 \over S_1 S_2 \dots
S_{n-5}}\label{npoint}, \eeqa
where the numerator is given as \beqa
\mathcal{N} = (135)(612)(235)(5~n-1~n)(13~n-1)\prod^{n}_{\alpha=8}(13\alpha)\prod^{n-1}_{\beta=7}(15 \beta)
\prod^{n-2}_{\gamma=6}(35 \gamma). \eeqa The sextics can be
written as
\begin{equation}
\begin{aligned}\label{nptsextics}
S_1&=(234)(456)(612)(135)-(123)(345)(561)(246),\\
S_2&=(n12)(13~n-1)(235)(5~n-1~n)-(123)(35~n-1)(5n2)(n-1~n~1),\\
S_{i-3}&=(i~i+1~i+2)(13~i+2)(15~i+1)(35i)-(135)(3i~i+2)(5i~i+1)(i+1~i+2~1),
\end{aligned}
\end{equation}
where $6\leq i\leq n-2$.

Several comments about this formula are in order.

Firstly, one can  deform the sextics by any
non-zero parameters $a_j$, namely 
\begin{equation}
S_{j}\rightarrow
S'_{j}=(klm)(mnp)(pqk)(qln)-a_j (qkl)(lmn)(npq)(kmp).
\label{deformation}
\end{equation}
As we will prove in next
section, interestingly, the final amplitude does not depend on $a_j$ at
all. Taking the limit  $a_j\rightarrow 0$ one gets ACCK
formula directly. This appears to be
a general fact, not specific to just NMHV amplitudes:
the ACCK Grassmanian integrand arises from the linked-connected
formula in a simple limit when the second terms in all
sextic polynomials are zero.

Secondly, the formula has $GL(3)$ symmetry for the Grassmanian, even
though we had started with the link representation for a particular
helicity configuration. We should point out that for some particular
gauge fixings, we do not always get the form of each sextic as a
polynomial of degree $6$ in the $c_{Ji} 's$. But, nevertheless, one
can numerically check that we do indeed get the tree amplitudes for
the connected prescription, namely, the residues at the locus
where all the sextics
simultaneously vanish.

Thirdly, writing sextics in terms of minors has a simple geometrical
interpretation\footnote{This was emphasised to us by Freddy
Cachazo.}. The minor $(i~j~k)=0$ in twistor space means the points
$i, j, k$ lie  on a line.  For NMHV, the sextics
$S_{ijk:lmn}=0$ means that these six points $i,j,k,l,m,n$ lie a conic
curve \cite{white}, which is consistent with the origin of the
connected prescription--integrating out degree two curves in twistor
space as in formula (\ref{eq:integralone}).

\section{From the Connected to ACCK  Using GRT}
In this section we will use the multidimensional Global Residue
Theorem (GRT) to analytically derive the BCFW contour of ACCK as in~(\ref{Nimabcf}) from the connected prescription formula
(\ref{npoint}).

 \subsection{n=6 and n=7}

We begin with $n=6$ and $n=7$ cases, which were previously done in \cite{Spradlin:2009qr}, \cite{Dolan:2009wf}.

$\bullet$ For the six-point amplitude, the connected formula gives
\begin{equation}
\begin{aligned}\label{A61}
A_6={(135) \over (123)(345)(561)} {1 \over S},
\end{aligned}
\end{equation}
where
\begin{equation}
S=(234)(456)(612)(135)-(123)(345)(561)(246).
\end{equation}

Cauchy's theorem states that the sum of residues in this expression is zero, so
\begin{equation}
\begin{aligned}\label{A62}
\{S\}=-\{1\}-\{3\}-\{5\},
\end{aligned}
\end{equation}
which is ACCK formula (\ref{Nimabcf}) for $n=6$.

$\bullet$
For the seven-point amplitude,
\beq
A_7=\frac{(135)(235)(612)(136)}{(123)(345)(671)}\frac{1}{S_1 S_2},
\label{A7}
\eeq
where
\beqa
 S_1  &=&  (234)(456)(612)(135) - (123)(345)(561)(246), \nn                     
 S_2  &=&  (567)(712)(235)(136) - (123)(356)(572)(671). 
                         \label{S7}
\eeqa
 By applying GRT, we get
\beqa
\{S_1,S_2\}=\{1,S_1\}+\{3,S_1\}+\{6,S_1\}.
\label{res7}
\eeqa
On the poles $(123)=0$ and $(345)=0$, the second term of $S_1$ vanishes and we get
\beq
\{1,S_1\}=\{1,2\}+\{1,4\},~~~~
 \{3,S_1\} = \cancel{\{3, 2\}} + \{3,4\}.
 \eeq
 Note that the terms with non-adjacent minors do not contribute because they would be cancelled by the numerator of $A_7$.
 Moreover, the condition of the residue $\{3,2\}$ implies that the points $2,3,4,5$ lie on a line and hence $(235)=0$, which is a term in the numerator of  $A_7$.
 To simplify  the residue $\{6,S_1\}$ we  use GRT again
\beqa
\{6,S_1\} &=& -\left( \{6,S_2\} + \{6,1\} +\{6,3\} \right)\\
                &=& -\left( \{6,5\} +\cancel{\{6,7\}}+ \{6,1\} +\{6,3\} \right).
\eeqa
Again,
$(671)=0$ makes the second term of $S_2$ vanish,
hence $\{6,S_2\}=\{6,5\}+\{6,7\}$. But the condition of $\{6,7\}$ implies that $(612)=0,$ which is a term in
the numerator of $A_7$. So finally, collecting all the residues we get \beqa
\{S_1,S_2\}=\{1,2\}+\{1,4\}+\{1,6\}+\{3,4\}+\{3,6\}+\{5,6\}.
\label{res71}
\eeqa
These are exactly the BCFW contours of the ACCK formula (\ref{Nimabcf}).

Let us conclude this section by saying that there are two useful properties which play an important role in making the above proof simple.
First, the second terms of the sextics vanish for some particular contours. Second, the residue vanishes if one of the non-adjacent minors
in the first term of the sextic vanishes.  We will use these two simple facts in the general proof, which follows in the next section.

\subsection{All $n$ proof }
Let us first note that one can easily check that the
second terms of the sextics vanish for any BCFW contours. It means that
whenever we get a BCFW contour~(\ref{Nimabcf}) by applying GRT, we
are sure that our NMHV formula for the amplitude is exactly of the same form as in ACCK amplitude, namely all
the non-adjacent minors cancel out.

We can further check that there are no `spurious' solutions, having non-vanishing contribution, from the connected contour.
Spurious solutions are those where the sextics vanish because individual minors in the expressions for the sextics vanish
(non-spurious solutions are those where the two terms in every sextic are separately non-zero).
We should exclude these solutions simply because the vanishing of any individual minor of the sextics means that the conic curve is not smooth anymore\footnote{The same reasoning holds for the validity of identity (\ref{identity}).}. 

The way to get BCFW contours from connected prescription is
simply to get rid of all the sextics in the connected contour by
applying GRT repeatedly. Let us remind you that
the poles in formula (\ref{npoint}) are
\begin{equation}
\begin{aligned}\label{pole}
(123)(345)(567)(n-1~ n~ 1)S_1S_{2}\dots S_{n-5}.
\end{aligned}
\end{equation}
Use GRT we have
\begin{equation}
\begin{aligned}\label{grt1}
\{S_{2}S_1\dots S_{n-5}\}=-&(\{1S_1S_3\dots S_{n-5}\}+\{3S_1S_3\dots
S_{n-5}\}\\
 &+\cancel{\{5S_1S_3\dots S_{n-5}\}}+\{(n-1)S_1S_3\dots S_{n-5}\} )\\
=-&(\{12S_3\dots S_{n-5}\}+\{14S_3\dots S_{n-5}\}+\{34S_3\dots
S_{n-5}\}\\ 
&+\cancel{\{32S_3\dots S_{n-5}\} }+\{(n-1)S_16\dots (n-2)\}),
\end{aligned}
\end{equation}
where $\{1S_1S_3\dots S_{n-5}\}$ is the residue of
$(123)=S_1=S_3=\dots=S_{n-5}=0$, and etc.

In order to explain why $\{5S_1S_3\dots S_{n-5}\}=0$  first notice
that $ \{5S_1S_3 S_4\dots S_{n-5}\}=\{5S_1678\dots (n-2)\}. $ This
is true because on $(567)=0$ the second term of $S_3$ vanishes and
hence $\{5S_1S_3\dots S_{n-5}\}=\{5S_16 S_4 \dots S_{n-5}\}$. Now in
addition to $(567)=0$, we also have $(678)=0$ which implies that the
points $5, 6, 7, 8$ lie on a line and hence $(578)=0$, resulting in the
second term of $S_4$ vanishing.
 So, we get $\{5S_16 S_4\dots S_{n-5}\}=\{5S_167 S_5 \dots S_{n-5}\}$. We can again apply similar arguments on $S_4$ and reduce it to $(789)$, and this goes on until the last sextic of the residue, which is $S_{n-5}$.
Now, $\{5S_1678\dots (n-2)\}$ means that the points $5, 6, \dots, n $
lie on a straight line, so $(5~n-1~n)$ in the numerator vanishes,
and hence $\{5S_1S_3\dots S_{n-5}\}=0$.

The equality  $\{(n-1)S_1S_3\dots S_{n-5}\} = \{(n-1)S_16\dots (n-2)\}$ in (\ref{grt1}) can also be explained along the same lines, but starting from the fact that, due to $(n-1~n1)=0$, $S_{n-5}$ is replaced by $(n-2~n-1~n)$. Finally $\{32S_3\dots S_{n-5}\}=0$ simply because $(345)=(234)=0$ implies $(235)=0$, which is a term in the numerator.

In the following, we will study each term from (\ref{grt1}) individually.  In the process, we will ignore all the vanishing terms without
explanation, since the reasons are very similar.

\subsubsection{$\{(n-1)S_16\dots
(n-2)\}$ term}

By applying
GRT again, with the poles
\beq
(123)(345)(567)(n-1~n1)S_1(n12)(678)(789) \dots (n-2~n-1~n),\nn
\eeq
we get the following non-vanishing residues
\begin{equation}
\begin{aligned}\label{i6}
-\{(n-1)S_16\dots
(n-2)\}=&\{(n-1)16\dots(n-2)\}+\{(n-1)36\dots(n-2)\}\\
&+\{(n-1)56\dots(n-2)\}.
\end{aligned}
\end{equation}
Actually these three terms are all the contours of the form $\{i6\dots\}$ and $i$ can be  $1,3$ or $5$, and they have the correct signs.

\subsubsection{$\{34S_3\dots S_{n-5}\}$ term}
 Now, in this case the poles are
 \beq
(123)(345)(567)(n-1~n1)(234)(456)(n12)S_3S_4 \dots S_{n-5}.\nn
\eeq
Again using GRT we get
\begin{equation}
\begin{aligned}\label{34I}
-\{34S_3\dots S_{n-5}\}=\underbrace{\{345S_4\dots
S_{n-5}\}}_{A_1}+\{34(n-1)7\dots(n-2)\}.
\end{aligned}
\end{equation}
The second term in the previous equation is a BCFW term and we use
GRT again on the term $A_1$ to generate another BCFW term in the
next step
\begin{equation}
\begin{aligned}\label{34II}
\{345S_4\dots S_{n-5}\}=-\Bigl(\underbrace{\{3456S_5\dots
S_{n-5}\}}_{A_2}+\{345(n-1)8\dots(n-2)\}\Bigr).
\end{aligned}
\end{equation}

Similarly, we can keep on using GRT repeatedly on one of the two
terms, generated at each step by using GRT in the previous step. In
the final step of this iteration, by applying GRT we get two terms,
$\{34567 \dots (n-4)(n-1)\}$ and $\{34567 \dots (n-3)\}$. So in this
way, we generate $\{347\dots (n-1)\}+\{3458\dots
(n-1)\}+\{34569\dots (n-1)\}+\dots+\{34567 \dots (n-3)\}$, which are
all the BCFW contours of the form $\{34\dots\}$.

\subsubsection{$\{14S_3\dots S_{n-5}\}$ term}
Now, let us consider the contours of the form $\{14 \dots \}$. Here
the poles are given as
\beqa
(123)(345)(567)(n-1~n1)(234)(456)(n12)S_3S_4\dots S_{n-5}. \eeqa
Using GRT we get the following \beqa
\begin{aligned}
-\{14S_3\dots S_{n-5}\} = \{14(n-1)7\dots(n-2)\} + \underbrace{\{142S_4\dots S_{n-5}\}}_{X_1}
                                     + \underbrace{\{145S_4\dots S_{n-5}\}}_{B_1}.
                                     \label{147I}
\end{aligned}
 \eeqa
Apart from the BCFW term $\{147\dots(n-1)\}$ we also have other
non-BCFW terms. Out of these, we will see that the terms like $X_1$
generated at each step will cancel out later from the same terms
generated by $\{12S_3 \dots S_{n-5}\}$ in the next subsection.
We can again apply GRT on $B_1$. Now, we can see the pattern of
BCFW terms generated from the $B_i$ terms, and here we will not
write the non-BCFW terms explicitly at each step
\begin{equation}
\begin{aligned}\label{147II}
&\{14S_3\dots S_{n-5}\}\Rightarrow \{147\dots(n-1)\},\\
&\{145S_4\dots S_{n-5}\}\Rightarrow \{1458\dots(n-1)\},\\
&\{1456S_5\dots S_{n-5}\}\Rightarrow \{14569\dots(n-1)\},\\
&\dots \dots
\end{aligned}
\end{equation}
In the final step of this series, by applying GRT, we have two
terms, $\{14567\dots(n-4)(n-1)\}$ and  $\{145678\dots(n-3)\}$. So by using GRT repeatedly, we get all the BCFW contours of
the type $\{14\dots \}$, namely
$\{147\dots(n-1)\}+\{1458\dots(n-1)\}+\{14569\dots(n-1)\}+\dots
+\{145678\dots(n-3)\}$.

\subsubsection{$\{12S_3\dots S_{n-5}\}$ term}
Finally, we look at the remaining contours $\{12S_3\dots S_{n-5}\}$ in equation
(\ref{grt1}).

Let us apply GRT and we get
\begin{equation}
\begin{aligned}\label{12361256}
-\{12S_3\dots S_{n-5}\}=&\{126\dots (n-3)3\}+\{126\dots
(n-3)5\}\\
&+\underbrace{\{12S_3\dots
S_{n-6}(n-1)\}}_{C_1}+\underbrace{\{12S_3\dots S_{n-6}4\}}_{D_1}.
\end{aligned}
\end{equation}
We can apply GRT on the term $C_1$ in (\ref{12361256}) again, and we
will deal with the term $D_1$ later. From $C_1$ we get \beqa
\begin{aligned}\label{123612561}
\{12S_3\dots S_{n-6}(n-1)\} = - &(\{126\dots (n-4)3(n-1)\} + \{126\dots
(n-4)5(n-1)\}\\
+&\underbrace{\{12S_3\dots S_{n-7}(n-2)(n-1)\}}_{C_2}+\underbrace{\{12S_3\dots S_{n-7}4(n-1)\}}_{E_1}) .
\end{aligned}
\eeqa We notice that one of the non-BCFW terms, $C_2$, is a similar kind of
term to $C_1$. Terms which are similar to $E_1$ and generated at
each step, will combine with other terms generated from the subsequent steps of
applying GRT.  The
general trend of BCFW contours generated from the $C_i$ terms are
\begin{equation}
\begin{aligned}\label{123612560}
&\{12S_3\dots S_{n-5}\} \Rightarrow \{1236\dots (n-3)\}+\{1256\dots
(n-3)\},\\
& \{12S_3\dots
S_{n-6}(n-1)\} \Rightarrow \{1236\dots (n-4)(n-1)\}+\{1256\dots
(n-4)(n-1)\},\\
& \{12S_3\dots
S_{n-7}(n-2)(n-1)\} \Rightarrow \{1236\dots (n-5)(n-2)(n-1)\}\\
&\hspace{2.3in}+\{1256\dots (n-5)(n-2)(n-1)\},\\
 &\dots \dots
\end{aligned}
\end{equation}
Note that at each step of the iteration we
also generate some non-BCFW terms(not explicitly written down in the
above pattern) which need to be dealt with as before. The final step in the above series generates the BCFW terms
$\{1238\dots (n-1)\}$,
$\{1258 \dots (n-1)\}$ and $\{1278\dots (n-1)\}$. %We should mention
%here that in the final step, apart from the BCFW terms we also
%generate some other terms which cancel out later.

By similar methods we can generate the other BCFW contours of the
form $\{12 \dots\}$ by using non-BCFW terms generated in previous
steps. Since all the steps are similar, here we only give some
examples of generating BCFW terms, without showing the details
\begin{equation}
\begin{aligned}\label{1234}
&\{12S_3\dots S_{n-6}4\}\Rightarrow \{12347\dots (n-3)\},\\
 &\{125S_4\dots
S_{n-6}4\}\Rightarrow \{123458\dots (n-3)\},\\
 &\{1256S_5\dots
S_{n-6}4\}\Rightarrow \{1234569\dots (n-3)\},\\
&\dots \dots
\end{aligned}
\end{equation}
Again the last step of this iterative process is special, the BCFW term
generated is $\{1234\dots (n-5)\}$. We will give a few examples of how non-BCFW terms combine to generate BCFW terms and we choose these particular examples as they give residues related to the ones in (\ref{1234}). Firstly
\begin{equation}
\begin{aligned}\label{12347}
&\{12(n-1)S_4\dots S_{n-6}4\}+\{12S_3\dots S_{n-7}4(n-1)\} \Rightarrow \{12347\dots (n-4)(n-1)\},\\
&\{12(n-1)S_4\dots S_{n-7}(n-2)4\}+\{12S_3\dots
S_{n-8}4(n-2)(n-1)\}\\ &\Rightarrow \{12347\dots
(n-5)(n-2)(n-1)\},\\
&\dots \dots
\end{aligned}
\end{equation}
The BCFW term generated from the last step of the above series is $\{12349\dots (n-1)\}$.  Next example  is
\begin{equation}
\begin{aligned}\label{123458}
&\{125(n-1)S_5\dots S_{n-6}4\}+\{12(n-1)S_4\dots S_{n-7}54\}\Rightarrow \{123458\dots (n-4)(n-1)\},\\
&\{125(n-1)S_5\dots S_{n-7}(n-2)4\}+\{12(n-1)S_4\dots
S_{n-8}5(n-2)4\}\\ &\Rightarrow \{123458\dots (n-5)(n-2)(n-1)\},\\
& \dots \dots
\end{aligned}
\end{equation}
The last step generates BCFW term $\{1234510\dots (n-1)\}$. And one more example will be
\begin{equation}
\begin{aligned}\label{1234569}
&\{125(n-1)S_5\dots S_{n-7}64\}+\{1256(n-1)S_6\dots S_{n-6}4\}\Rightarrow \{1234569\dots (n-4)(n-1)\},\\
&\{125(n-1)S_5\dots S_{n-8}6(n-2)4\}+\{1256(n-1)S_6\dots
S_{n-7}(n-2)4\}\\
&\Rightarrow \{1234569\dots (n-5)(n-2)(n-1)\}, \\
&\dots \dots
\end{aligned}
\end{equation}
The BCFW term generated in the last step is $\{12345611\dots (n-1)\}$.

From the above mentioned examples, we can see the general pattern: the first term in
(\ref{1234}), $\{12347\dots (n-3)\}$, combining with all the terms
from (\ref{12347}) generates all the contours of the form
$\{12347\dots \}$; similarly, the second term in (\ref{1234}),
$\{123458\dots (n-3)\}$,  and all the terms in (\ref{123458}) give us all
the contours of the form $\{123458 \dots\}$; the third term in
(\ref{1234}), $\{1234569\dots (n-3)\}$, together with all the terms of
(\ref{1234569}) give us all the contours of the form $\{1234569
\dots\}$. It is not hard to see that all the other BCFW  terms of
the form $\{1234\dots \}$ can be generated in a similar way. So we
have generated all the contours of the form $\{12 \dots\}$ and we notice that they can be grouped into contours of the form, $\{1236 \dots\}$, $\{1256 \dots\}$, $\{1238
\dots\}$, $\{1258 \dots\}$, $\{1278 \dots\}$ and $\{1234 \dots\}$.

As we had seen so far, each GRT step also generates terms which have no
contribution to BCFW contours. These terms, typically, look like
$\{124\dots i, S_i, \dots, S_{n-5}\}$, but they just cancel out at
each step. At each of the final steps, we also generate terms like
$\{124\dots i, i+4, \dots, (n-1)\}$ and $\{124\dots(n-4)\}$, and
they also cancel out. In the Appendix we can see all these
cancelations explicitly in the $10$-point example.

Let us conclude with our main result
\begin{equation}
\begin{aligned}\label{PBCFW}
\oint_C \frac{\mathcal{N}}{(123)(345)(567)(n-1~n~1)} {1 \over S_1
S_2 \dots S_{n-5}}= \oint_{B} {1 \over (123)(234) \dots (n12)},
\end{aligned}
\end{equation}
where contour $C$ is the connected contour, and $B$ is the BCFW
contour. One can apply GRT again and show that the same equality is true for
the P(BCFW) contour.

Since for any BCFW contour the second terms of sextics vanish, so as a byproduct, we also proved the statement we made before that deforming sextics by some non-zero parameters still gives us the correct tree amplitude.

\section*{Acknowledgments}

We are grateful to N.~Arkani-Hamed, F.~Cachazo, L.~Dolan, R.~Roiban, D.~Skinner, C.~Vergu and
especially M.~Spradlin for very useful conversations. 
This work was
supported in part by the US Department of Energy under contract
DE-FG02-91ER40688
 and the US National Science Foundation under grants
PECASE PHY-0643150 PHY-0548311.

\appendix
\section{Linked-Connected to BCFW--10 Point Amplitude}
Let us consider one higher-point non-trivial case, the $10$-point amplitude, here we will ignore all vanishing terms. For $n=10$,
the poles are $(123)(345)(567)(9~10~1)S_1S_2S_3S_4S_5$. Using GRT, we have
\begin{equation}
\begin{aligned}\label{grt10}
\{S_{2}S_1S_3S_4S_{5}\}&=-\left(\{1S_1S_3S_4 S_{5}\}+\{3S_1S_3S_4
S_5\}+\{5S_1S_3S_4
S_5\}+\{9S_1S_3S_4S_5\} \right)\\
&=-\left(\{12S_3S_4S_{5}\}+\{14S_3S_4S_{5}\}+\{34S_3S_4
S_{5}\}+\{9S_1678\} \right).
\end{aligned}
\end{equation}
As in the general case, we can apply GRT again
\begin{equation}
\begin{aligned}\label{1610}
-\{9S_1678\}=\{91678\}+\{93678\}+\{95678\},
\end{aligned}
\end{equation}
and
\begin{equation}
\begin{aligned}\label{3410}
-\{34S_3S_4
S_{5}\}=&\{345S_4S_5\}+\{34978\}\\=&-\left(\{3456S_5\}+\{34598\}\right)+\{34978\}\\
=&\{34569\}+\{34567\}+\{34589\}+\{34789\}.
\end{aligned}
\end{equation}
Now let us consider the contours of the form $\{12 \dots \}$ and $\{14 \dots\}$.
\begin{equation}
\begin{aligned}\label{121}
-\{12S_3S_4
S_{5}\}=&\{12S_3S_43\}+\{12S_3S_45\}+\{12S_3S_49\}+\{12S_3S_44\}\\
=&\{12673\}+\{12675\}+\{12S_3S_49\}+\{12S_3S_44\}.
\end{aligned}
\end{equation}
Applying GRT again we get
\begin{equation}
\begin{aligned}\label{122}
\{12S_3S_49\}=&-\left(\{12639\}+\{12659\}+\{12S_349\}+\{12S_389\}
\right),\\
\{12S_3S_44\}=&-\Bigl(\{12374\}+\{125S_44\}+\{129S_44\}+\underbrace{\{12S_5S_44\}}_{1}
\Bigr),
\end{aligned}
\end{equation}
and
\begin{equation}
\begin{aligned}\label{141}
-\{14S_3S_4
S_{5}\}=&\{145S_4S_5\}+\{14978\}+\{142S_4S_5\}\\
=&-\left(\{14598\}+\{1452S_5\}+\{1456S_5\}\right)+\{14978\}+\{142S_4S_5\}\\
=&\{14589\}+\{14978\}+\underbrace{\{142S_4S_5\}}_{1}-\underbrace{\{1452S_5\}}_{2}
+\{14569\}+\underbrace{\{14562\}}_{3}+\{14567\},
\end{aligned}
\end{equation}
and also
\begin{equation}
\begin{aligned}\label{123}
-\{12S_389\}=&\{12389\}+\{12589\}+\underbrace{\{12489\}}_{4}+\{12789\},\\
-\{125S_44\}=&\{12534\}+\underbrace{\{12594\}}_{5}+\underbrace{\{12564\}}_{3}+\underbrace{\{125S_54\}}_{2}.
\end{aligned}
\end{equation}
Now we take $\{12S_349\}$ and $\{129S_44\}$ from equation (\ref{122}) and
applying GRT we get
\begin{equation}
\begin{aligned}\label{124}
\{129S_44\}+\{12S_349\}=\{12934\}+\underbrace{\{12954\}}_{5}+\underbrace{\{12984\}}_{4}.
\end{aligned}
\end{equation}
Note the all the underbraced terms cancel out \footnote{The numbering marks out which terms cancel.}, and it is a
general feature even for general $n$-point cases as we pointed out
before. Now we collect all the non-vanishing term together
\begin{equation}
\begin{aligned}\label{tree10}
\{S_2S_1S_3S_4S_5\}=&\{91678\}+\{93678\}+\{95678\}+\{34569\}+\{34567\}+\{34589\}+\{34789\}
\\+&\{12673\}+\{12675\}-\{12639\}-\{12659\}-\{12374\}+\{14589\}+\{14978\}\\
+&\{14569\}+\{14567\}+\{12389\}+\{12589\}+\{12789\}+\{12534\}+\{12934\},
\end{aligned}
\end{equation}
which are exactly the BCFW contours for $10$ points as predicted by the
ACCK formula (\ref{Nimabcf}), and have the right signs.

\end{document}